\documentclass[10pt,prd,aps,onecolumn,nofootinbib,superscriptaddress]{revtex4-2}
\usepackage{amsmath}
\usepackage{booktabs}
\usepackage{graphicx}
\usepackage[dvipsnames,table]{xcolor}
\usepackage[colorlinks=True, linkcolor=MidnightBlue, citecolor=ForestGreen]{hyperref}
\usepackage[range-phrase=\textendash, range-units=single]{siunitx}
\DeclareSIUnit\day{day}
\DeclareSIUnit\pc{pc}

\begin{document}

\title{Lorentz-Invariance Violation with neutrinos from gamma-ray bursts\texorpdfstring{\\}{ }using IceCube High-Energy Starting Events 12-year dataset}
\author{Mathieu Lamoureux}
\email{lamoureux@apc.in2p3.fr}
\affiliation{Université Paris Cité, CNRS, Astroparticule et Cosmologie, F-75013 Paris, France} 
\author{José Manuel Carmona}
\email{jcarmona@unizar.es}
\affiliation{Centro de Astropartículas y Física de Altas Energías (CAPA), Universidad de Zaragoza}
\affiliation{Departamento de Física Teórica, Universidad de Zaragoza, E-50009 Zaragoza, Spain}
\author{José Luis Cortés}
\email{cortes@unizar.es}
\affiliation{Centro de Astropartículas y Física de Altas Energías (CAPA), Universidad de Zaragoza}
\affiliation{Departamento de Física Teórica, Universidad de Zaragoza, E-50009 Zaragoza, Spain}
\author{Ardit Gkioni}
\email{agkioni@unizar.es}
\affiliation{Centro de Astropartículas y Física de Altas Energías (CAPA), Universidad de Zaragoza}
\affiliation{Departamento de Física Teórica, Universidad de Zaragoza, E-50009 Zaragoza, Spain}
\author{Antoine Kouchner}
\affiliation{Université Paris Cité, CNRS, Astroparticule et Cosmologie, F-75013 Paris, France} 
\author{Maykoll A. Reyes}
\email{mkreyes@unizar.es}
\affiliation{University of Rijeka, Faculty of Physics, Rijeka 51000, Croatia}
\affiliation{Centro de Astropartículas y Física de Altas Energías (CAPA), Universidad de Zaragoza}
\date{October 2026}

\begin{abstract}
High-energy astrophysical neutrinos provide a unique probe of Lorentz-invariance violation (LIV), as small energy-dependent deviations from the speed of light can produce measurable time delays over cosmological distances. We search for LIV-induced correlations between neutrinos and gamma-ray bursts using the public 12-year IceCube High-Energy Starting Event dataset and a catalog of 746 bursts with measured redshifts. We develop a likelihood analysis accounting for neutrino directional and energy uncertainties, a possible intrinsic time offset between neutrino and gamma-ray emission, and the decay of superluminal neutrinos expected within effective field theory. For linear LIV, the data show a mild preference over the background-only hypothesis, with a $p$-value of $3.1\%$, corresponding to a one-sided Gaussian significance of approximately $1.9\sigma$. The best fit corresponds to a LIV scale $\Lambda_{\rm LIV}=5.8\times10^{17}\,\mathrm{GeV}$ and is effectively driven by the subluminal component, as superluminal neutrinos are strongly suppressed by LIV-induced decays at the analyzed energies. The preferred LIV scale and the most significant neutrino--gamma-ray burst associations are consistent with previous studies based on different statistical approaches. No evidence for quadratic LIV is found. These results motivate further searches with complementary neutrino datasets, particularly at lower energies, where the superluminal component may become observable.
\end{abstract}

\maketitle

\section{Introduction}
\label{sec:intro}

High-energy astrophysical neutrinos provide a unique opportunity to probe fundamental physics over energy and distance scales inaccessible in terrestrial experiments, as displayed by the recent detection of the KM3-230213A event~\cite{KM3NeT:2025npi}. Unlike photons, neutrinos can travel essentially unimpeded across cosmological distances, preserving information about both their production mechanisms and their propagation. For this reason, neutrino astronomy has become an important laboratory to test possible deviations from standard physics, including violations of Lorentz invariance (LIV), a fundamental symmetry of the Standard Model (SM) of particle physics.

LIV is a generic prediction of several approaches to quantum gravity~\cite{Ellis:2008gg,Bojowald:2004bb,Horava:2009uw,Loll:2019rdj,Carlip:2017eud}. Within effective field theory, these effects can be described by the Standard Model Extension (SME), where Standard Model fields couple to preferred background tensors. The minimal SME~\cite{Colladay:1998fq} contains renormalizable operators leading to energy-independent LIV effects, whereas non-renormalizable operators involving additional derivatives~\cite{Myers:2003fd} generate energy-dependent modifications of the particle dispersion relations. The latter are particularly relevant for astrophysical neutrinos, since their effects grow with energy and accumulate over cosmological propagation distances. In particular, if neutrino velocities deviate slightly from the speed of light, neutrinos emitted simultaneously with photons could arrive earlier or later depending on their energy. Searches for such time-of-flight effects have therefore focused on transient astrophysical phenomena, where a well-defined emission time can be identified.

Gamma-ray bursts (GRBs) are especially attractive targets in this context. They are among the most energetic explosions in the Universe, are observed up to cosmological distances, and provide precise timing information through their gamma-ray emission. Moreover, GRBs were historically considered promising sources of high-energy neutrinos~\cite{Waxman:1997ti}, motivating extensive searches for temporal and directional coincidences between GRBs and neutrino events~\cite{IceCube:2012qza,ANTARES:2013ath}.

Paradoxically, despite more than a decade of observations by IceCube and other neutrino telescopes, no statistically compelling population of neutrinos associated with GRBs has been established~\cite{ANTARES:2016fmg,Abbasi:2022whi,IceCube:2023rhf}. This absence of correlations remains one of the outstanding puzzles of multimessenger astrophysics. While it may indicate that GRBs are not the dominant sources of the observed diffuse neutrino flux, it also raises the possibility that neutrino propagation differs from the standard expectation. In particular, if neutrinos experience energy-dependent propagation effects, genuine source associations could be obscured by arrival-time shifts accumulated over astrophysical distances.

In this work, we perform a systematic search for LIV signatures in neutrino--GRB correlations using publicly available datasets. Unlike previous approaches~\cite{Amelino-Camelia:2016ohi, Huang:2018ham, Huang:2019etr,Amelino-Camelia:2022pja,Amelino-Camelia:2025lqn,Wang:2026eyt}, our analysis is based on a statistical likelihood framework that accounts for both a possible intrinsic time delay and a deformation of the neutrino spectrum induced by the decay of superluminal neutrinos. A positive observation would have profound implications for fundamental physics, potentially revealing departures from Lorentz invariance. Otherwise, the absence of statistically significant correlations can be used to constrain the LIV effects and the corresponding LIV scale.

The structure of the paper is as follows: \autoref{sec:theory} summarizes the theoretical framework and notations used throughout the article; the inputs and the statistical approach are introduced in \autoref{sec:method}; the results are presented and discussed in \autoref{sec:results} and \autoref{sec:discussion}, respectively.

\section{Theoretical framework}
\label{sec:theory}

We are interested in LIV effects that may induce an energy-dependent propagation velocity for very high-energy astrophysical neutrinos, for which we will neglect their masses. For corrections of order $n\geq1$, the neutrino group velocity can be parametrized as
\begin{equation}
   v \simeq
1\pm\frac{n+1}{2}\eta^n
\left(\frac{E_\nu}{M_{\rm P}}\right)^n,
\label{eq:velocity_nu}
\end{equation} 
where the positive (negative) sign corresponds to superluminal (subluminal) propagation, $M_{\rm P}=1.22\times10^{19}\,\mathrm{GeV}$ is the Planck mass, and
\begin{equation}
    \eta \equiv \frac{M_{\rm P}}{\Lambda_{\rm LIV}}
    \label{eq:eta}
\end{equation}
parametrizes the strength of the LIV effect, with $\Lambda_{\rm LIV}$
the characteristic energy scale suppressing the LIV correction. The Lorentz-invariant limit is recovered for $\eta=0$, corresponding to the $\Lambda_{\rm LIV}\to\infty$ limit.

Within effective field theory, this phenomenological parametrization  can arise from a rotationally invariant LIV correction to the neutrino kinetic term,
\begin{equation}
\mathcal{L}_{\rm LIV}^{(\nu)}
=+\frac{\eta^n}{2M_{\rm P}^n}
\bar{\nu}_L\gamma^0(i\partial_0)^{n+1}\nu_L.
\label{eq:LIV_lagrangian}
\end{equation}
The corresponding modified Dirac equation leads to different dispersion relations for neutrinos and antineutrinos~\cite{Carmona:2022dtp}. In the ultrarelativistic limit,
\begin{equation}
E_\nu \simeq p
-\frac{1}{2}\eta^n\frac{p^{n+1}}{M_{\rm P}^n},
\end{equation}
while
\begin{equation}
E_{\bar{\nu}} \simeq p
-(-1)^n\frac{1}{2}\eta^n\frac{p^{n+1}}{M_{\rm P}^n}.
\end{equation}
As a consequence, neutrinos and antineutrinos acquire LIV corrections with opposite signs for odd $n$, and with the same sign for even $n$. In particular, for linear ($n=1$) LIV, one component is necessarily superluminal while the other is subluminal. With the sign convention adopted above, neutrinos are subluminal, and antineutrinos are superluminal. The opposite assignment would result from reversing the sign of the Lagrangian correction in \autoref{eq:LIV_lagrangian} and is phenomenologically equivalent for the present analysis, which does not distinguish neutrinos from antineutrinos. For quadratic ($n=2$) LIV, the same convention corresponds to both neutrinos and antineutrinos being subluminal, which is the scenario considered below.

\subsection{Time-of-flight effects}

An energy-dependent neutrino velocity leads to an accumulated difference between the arrival times of neutrinos and photons emitted by the same astrophysical source. For a neutrino with energy $E_\nu$ at Earth from a source at redshift $z$, the LIV-induced contribution to the time delay between this neutrino and a photon is
\begin{equation}
    \Delta t_{\rm LIV}=\pm \frac{n+1}{2H_0}
    \left(\frac{\eta E_\nu}{M_{\rm P}}\right)^n
    \int_0^z \frac{(1+z')^n}
    {\sqrt{\Omega_m(1+z')^3+\Omega_\Lambda}} \,{\rm d}z',
\end{equation}
where the positive (negative) sign corresponds to subluminal (superluminal) propagation, $H_0$ is the Hubble constant, and $\Omega_m$ and $\Omega_\Lambda$ denote the present-day matter and dark-energy density parameters, respectively. In the following, values of $H_0 = \qty{67.4}{\km\per\s\per\mega\pc}$, $\Omega_m = 0.315$, and $\Omega_\Lambda = 0.685$ are taken from Planck measurements~\cite{Planck:2018vyg}. It is worth noting that the discrepancy between different measurements of the Hubble constant (known as Hubble tension~\cite{DiValentino:2021izs}) would directly translate into up to 10\% variation of the time delay and thus an impact on the LIV scale value of the same order (for $n=1$). It is convenient to introduce
\begin{equation}
    g_n(z) = \frac{n+1}{2H_0M_{\rm P}^n}\int_0^z \frac{(1+z')^n}
    {\sqrt{\Omega_m(1+z')^3+\Omega_\Lambda}} \,{\rm d}z',
    \label{eq:g_ndef}
\end{equation}
so that the propagation-induced time delay takes the form
\begin{equation}
    \Delta t_{\rm LIV} = \pm(\eta E_\nu)^n g_n(z).
    \label{eq:time-delay-liv}
\end{equation}
For $n=1$ and $z=1$, we get
\begin{equation}
    \left.\Delta t_{\rm LIV}\right\vert_{n=1,z=1} \simeq \pm \eta \times \left(\dfrac{E_\nu}{\qty{100}{\TeV}}\right) \times \qty{4.8e-2}{\day}.
\end{equation}

In addition to this propagation effect, neutrinos and gamma rays may not be emitted simultaneously at the source. Denoting their intrinsic source-frame time difference by $\Delta t_{\rm in}$, the observed time difference is therefore
\begin{equation}
    \Delta t_{\rm obs} = t_\nu-t_{\rm GRB} =
    (1+z)\Delta t_{\rm in}
    \pm(\eta E_\nu)^n g_n(z),
    \label{eq:time-delay}
\end{equation}
where the first term accounts for cosmological time dilation of the intrinsic emission offset, while the second describes the LIV contribution accumulated during propagation.

\subsection{Decay of superluminal neutrinos}

Besides modifying their time of flight, LIV can render sufficiently energetic superluminal neutrinos unstable. The dominant decay channels are vacuum pair emission (VPE),
\begin{equation}
    \nu_\alpha \rightarrow \nu_\alpha+e^-+e^+,
\end{equation}
and neutrino splitting (NSpl),
\begin{equation}
    \nu_\alpha \rightarrow
    \nu_\alpha+\nu_\beta+\bar{\nu}_\beta.
\end{equation}
The VPE process is only kinematically allowed above the threshold
\begin{equation}
    E_{\rm th}^{\rm VPE}=\left(
    \frac{4m_e^2M_{\rm P}^n}{\eta^n}
    \right)^{1/(n+2)}.
\end{equation}
The threshold for neutrino splitting is negligible at the energies and LIV scales relevant here because of the smallness of the neutrino masses.

Far above threshold, the decay widths exhibit a strong dependence on both energy and the LIV scale. For a process $A\in\{\mathrm{VPE},\mathrm{NSpl}\}$, they can be written as
\begin{equation}
    \Gamma_{\alpha}^{(A)}(E)=
    K_{\alpha}^{(A)}
    \frac{G_F^2}{192\pi^3}
    \frac{\eta^{3n}E^{5+3n}}{M_{\rm P}^{3n}},
    \label{eq:asymptotic-decay-width}
\end{equation}
where $G_F$ is the Fermi constant and $K_{\alpha}^{(A)}$ is a dimensionless coefficient that depends on the decay channel, neutrino flavour, and order $n$ of the LIV correction~\cite{Carmona:2022dtp,Carmona:2025lir}. The VPE contribution is set to zero below $E_{\rm th}^{\rm VPE}$. The total decay width is thus
\begin{equation}
    \Gamma_\alpha^{\rm tot}(E)=
    \Gamma_\alpha^{\rm NSpl}(E)+
    \Gamma_\alpha^{\rm VPE}(E)\,\,\mathbf{H}\!\left(E-E_{\rm th}^{\rm VPE}\right),
\end{equation}
where $\mathbf{H}$ denotes the Heaviside function.

During propagation from cosmological distances, the neutrino energy evolves as $E(z')=(1+z')E_\nu$, where $E_\nu$ is the energy at Earth. The probability for a superluminal neutrino emitted at redshift $z$ to reach Earth without decaying is consequently
\begin{equation}
    \mathcal{P}_{\rm surv}(E_\nu,\eta,z)=
    \exp\left[-\tau(E_\nu,\eta,z)\right],
    \label{eq:survival_probability}
\end{equation}
where the optical depth is
\begin{equation}
    \tau(E_\nu,\eta,z)=\int_0^z
    \frac{\Gamma_\alpha^{\rm tot}\left((1+z')E_\nu\right)}{(1+z')H_0\sqrt{\Omega_m(1+z')^3+\Omega_\Lambda}}\,{\rm d}z'.
    \label{eq:optical_depth}
\end{equation}
Because of the strong energy dependence of the decay width, the survival probability rapidly decreases above a given energy scale~\cite{Carmona:2023mzs}. Superluminal neutrinos at sufficiently high energies are therefore strongly suppressed after propagation over cosmological distances.

In the following analysis, we will first consider the $n=1$ case, for which neutrinos and antineutrinos acquire LIV corrections of opposite sign. An astrophysical flux consequently contains both a subluminal component, arriving later than in the Lorentz-invariant case, and a superluminal component, arriving earlier and additionally suppressed according to \autoref{eq:survival_probability}. The resulting signal is thus asymmetric between positive and negative time delays. These propagation effects are incorporated into the signal time probability density introduced in the following section.

We will also present an analysis for the $n=2$ case in the subluminal scenario, for which no LIV-induced decay is present. The superluminal scenario is not phenomenologically viable in the parameter region relevant to this type of time-delay analysis, since both neutrinos and antineutrinos would be unstable and, given the steep energy dependence of the decay rates, values of the LIV parameter large enough to produce observable time delays would be incompatible with the observation of astrophysical neutrinos at PeV energies~\cite{Carmona:2023mzs}.

\section{Method}
\label{sec:method}

The analysis relies on three main ingredients: the selection and characterization of the neutrino dataset, the GRB catalog, and the statistical analysis used to quantify potential correlations.

\subsection{Neutrino dataset}
\label{sec:nu-dataset}

As the expected time delay depends on the neutrino energy, an ideal sample would consist of well-localized very-high-energy neutrinos with relatively small energy uncertainties (since these directly translate into uncertainties in their time delay). The High-Energy Starting Event (HESE) dataset is therefore perfectly suited for the search, as it contains events with reconstructed energies from \qty{20}{\tera\eV} to \qty{5}{\peta\eV}, with direction uncertainties of up to a few degrees. The 12-year data release~\cite{DVN/PZNO2T_2023} contains $N_{\nu} = 164$ events, including 55 classified as tracks and 109 classified as cascades, detected between 2010 and 2022.

For each event, the data release provides the direction and its uncertainty using the 8-parameter Fisher-Bingham probability distribution $p_{\text{FB}}(\Omega)$. The time (in Modified Julian Date) and the reconstructed energy  $E_{\rm rec}$ (in GeV) are also provided. However, there is no event-by-event uncertainty on the latter. In the following, a Gaussian energy resolution of $\varepsilon_E = 30\%$ ($11\%$) is assumed for tracks (cascades), as described in \cite{IceCube:2020wum}: 
\begin{equation}
    \dfrac{{\rm d}P}{{\rm d}E_\nu}(E_\nu, E_{\rm rec}) = \dfrac{1}{\varepsilon_E E_{\rm rec} \sqrt{2\pi}} \times \exp \left[ - \dfrac{(E_\nu - E_{\rm rec})^2}{2(\varepsilon_E E_{\rm rec})^2} \right] \times \mathbf{H}(E_\nu).
    \label{eq:energy-pdf}
\end{equation}
where the Heaviside function ensures that only positive energies are considered.

\subsection{Gamma-ray burst catalog and modeling}
\label{sec:grb-catalog}

The list of GRBs is extracted from the GRBweb database~\cite{GRBweb}. Out of the more than 9,000 GRBs that occurred between 1997 (start of the collection) and May 2026 (when the analysis was performed), only those with a measured redshift are retained, reducing the catalog to $N_{\rm GRBs} = 746$ bursts. The rate of selected GRBs is higher after 2005 (following the start of the Swift mission).

Rather than assuming a specific physical model for neutrino production in GRBs, we adopt a simple phenomenological prescription for the emission. The intrinsic time delay between the neutrino and the gamma-ray emission is assumed to be uniformly distributed in a short time window,
\begin{equation}
    \Delta t_{\rm in} = t_\nu - t_{\rm GRB} \in [\tau_-, \tau_+] = [-350, +150]\,\unit{\s},
\end{equation}
following Ref.~\cite{Baret:2011tk}. A single power law $E^{-2}$ is considered for the neutrino spectrum.

Furthermore, as the search is performed over a large population of GRBs rather than on individual ones, weighting is required to assess the contribution of each source. Though a distance-dependent weight (corrected by the source luminosity distance) may be more appropriate, as this leads to a few GRBs dominating (the closest ones), an equal weighting $w_j = 1/N_{\rm GRBs}$ is assumed.

\subsection{Statistical analysis}
\label{sec:stat}

The goal of the search is to identify the eventual presence of LIV effects in the association between neutrinos and GRBs. Therefore, there are two parameters of interest: $\eta$, defined in \autoref{eq:eta}, and the fraction of neutrinos from GRBs in the neutrino dataset $f_{\rm sig}$ ($f_{\rm sig} \in [0, 1]$). In the absence of LIV, $\eta = 0$, while $f_{\rm sig} = 0$ is an indication that no neutrino in the dataset is associated with a GRB (whether LIV exists or not).

The analysis is performed using a standard point-source likelihood, defined as
\begin{equation}
    \mathcal{L}(f_{\rm sig}, \eta) = \prod_{i=1}^{N_\nu} \left[f_{\rm sig} \sum_{j=1}^{N_{\rm GRBs}} w_j \mathcal{S}_{i,j}(\eta) + \left(1-f_{\rm sig}\right) \mathcal{B}_i \right]
    \label{eq:lkl}
\end{equation}
where the product runs over neutrino events in the dataset, the inner sum runs over all GRBs in the catalog, $S_{i,j}$ is the signal PDF term for event $i$ and source $j$, and $B_i$ is the background PDF term for event $i$.

Both signal and background PDFs are defined as the product of an angular term and a time term, $\mathcal{S} = \mathcal{S}^{\rm ang} \times \mathcal{S}^{\rm time}$ and $\mathcal{B} = \mathcal{B}^{\rm ang} \times \mathcal{B}^{\rm time}$. The background angular term is derived using data and depends solely on the declination: $B_{i}^{\rm ang} = f(\delta)$. The background time term is assumed to be uniform between the first and the last event in the dataset: $B_{i}^{\rm time} = 1/(t_{\max} - t_{\min})$. The signal angular term is directly computed following the Fisher-Bingham parametrisation in the dataset: $S_{i,j}^{\rm ang} = p_{\text{FB},i}(\Omega_j)$.

\begin{figure}[hbtp]
    \centering
    \includegraphics[height=0.37\linewidth]{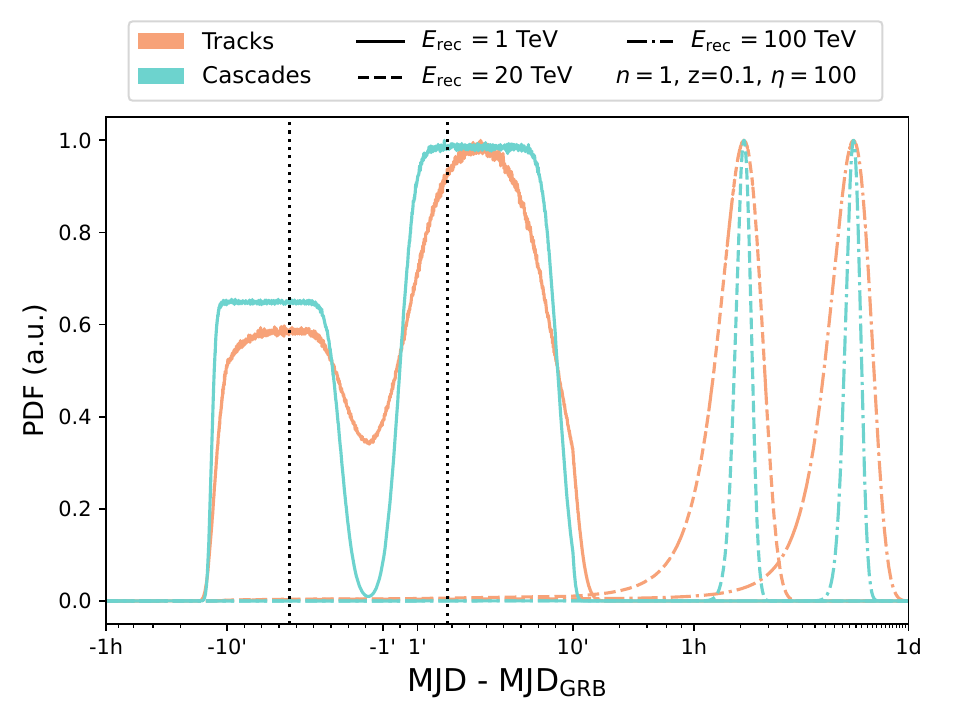}
    \includegraphics[height=0.37\linewidth]{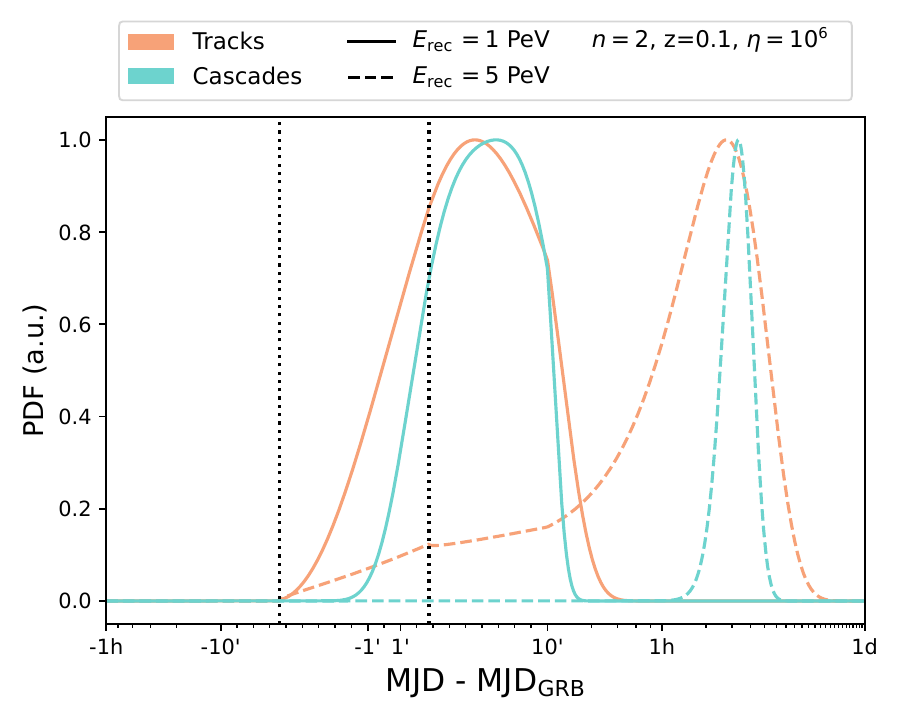}
    \caption{Signal time PDF as a function of the time difference $\Delta t$ between the neutrino and the GRB onset, for first-order LIV with $\eta = 100$ (left) and second-order LIV with $\eta = \num{e6}$ (right), assuming a source at redshift $z=0.1$. The orange (blue) line corresponds to track (cascade) events, and the different line styles indicate the reconstructed neutrino energy. The horizontal axis uses a linear scale for $|\Delta t|\leq 10$ min and a logarithmic scale beyond this range. The dotted vertical lines show the intrinsic time window $[-(1+z) \times 350, (1+z) \times 150]$\,s. Note that the case $E_{\rm rec} = \qty{1}{\TeV}$ is not representative of the IceCube HESE dataset, which has only events at higher energies, but is used to illustrate the contribution of superluminal neutrinos to the left-hand branch of the distribution.}
    \label{fig:time-pdf}
\end{figure}

The signal time term is the convolution of three effects: the intrinsic time delay between the neutrino and the gamma-ray signals, the additional time delay due to LIV, and the deformation linked to the decay of the superluminal component. For a LIV correction of order $n$, the time delay is written as in \autoref{eq:time-delay},
\begin{equation}
    \Delta t_{i,j} = t_i - t_j = (1+z_j) \Delta t_{\rm in} \pm (\eta E_{\nu})^n \times g_n(z_j),
    \label{eq:time-delay2}
\end{equation}
where the plus (minus) sign corresponds to subluminal (superluminal) neutrinos. As the energy in \autoref{eq:time-delay2} is the neutrino energy at Earth, one should account for the uncertainty on its value based on the reconstructed energy. For $n = 1$, the signal time term is then
\begin{align}
    S_{i,j}^{\rm time} &= \dfrac{{\rm d}P}{{\rm d}t_i}(t_i; E_{\rm rec, i}, \eta, z_j) \nonumber \\
    &= C \times \left[ \int_{E_{+}}^{E_{-}} \dfrac{{\rm d}P}{{\rm d}E_\nu} \left(E_\nu, E_{\rm rec, i} \right) {\rm d}E_\nu + \int_{-E_{-}}^{-E_{+}} \dfrac{{\rm d}P}{{\rm d}E_\nu} \left(E_\nu, E_{\rm rec, i} \right) \times \mathcal{P}_{\rm surv} \left(E_\nu, \eta, z_j \right) {\rm d}E_\nu \right]
    \label{eq:time-pdf}
\end{align}
where the first (second) term is the subluminal (superluminal) component, $E_{\pm}(\eta, t_i, t_j, z_j) = [\eta g_1(z_j)]^{-1} \times [t_i - t_j - (1 + z_j)\tau_\pm]$, ${\rm d}P/{\rm d}E_\nu(E_\nu, E_{\rm rec, i})$ is the neutrino energy PDF described in \autoref{sec:nu-dataset}, $\mathcal{P}_{\rm surv}(E_\nu, \eta, z)$ is the survival probability function defined in \autoref{eq:survival_probability}, and $C$ is a normalization ensuring that $\int {\rm d}P/{\rm d}t_i(t_i; \ldots) {\rm d}t_i = 1$. An illustration of the time PDF is shown in the left panel of \autoref{fig:time-pdf} for $\eta=50$.

For second-order LIV ($n=2$), both neutrinos and antineutrinos are assumed to be subluminal and, therefore, only the first term appears in \autoref{eq:time-pdf}, with $E_{\pm} = [\eta^2 g_2(z_j)]^{-1/2} \times [t_i - t_j - (1 + z_j)\tau_\pm]^{1/2}$ and $C = 1$. An example is shown in \autoref{fig:time-pdf}.

The test statistic is defined as the log-likelihood ratio between the best fit and the null hypothesis ($f_{\rm sig}=0$ and $\eta=0$):
\begin{equation}
    \mathcal{T} = 2 \log \dfrac{\mathcal{L}(\hat{f}_{\rm sig}, \hat{\eta})}{\mathcal{L}(0, 0)}
\end{equation}
where the hatted values are the values maximizing the likelihood, computed using a scan in $\eta$ in the range $[\num{e-4}, \num{e4}]$ ($[\num{e4}, \num{e14}]$) for $n = 1$ ($n = 2$), and a binary search of the root of ${\rm d}\mathcal{L}/{\rm d}f_{\rm sig}$ in the range $f_{\rm sig} \in [0, 1]$.

Background-only pseudo-experiments are generated using scrambled data, where the times of neutrino events are shuffled, their right ascensions are recomputed accordingly, and the times and right ascensions of GRBs are also shuffled. From these pseudo-experiments, the probability distribution $f_{\text{bkg}}(\mathcal{T})$ is estimated, and the final p-value $p$ of the observation is computed as:
\begin{equation}
    p = \int_{\mathcal{T}_{\rm obs}}^\infty f_{\text{bkg}}(\mathcal{T}) d\mathcal{T}
\end{equation}
where $\mathcal{T}_{\rm obs}$ is the value of the test statistic for the data without any scrambling. We can also extract the best-fit values $\hat{f}_{\rm sig}$ and $\hat{\eta}$ as well as likelihood contours, applying Wilks' theorem on the log-likelihood ratio $\Delta\chi^2 = 2 \times \left[ \log \mathcal{L}(\hat{f}_{\rm sig}, \hat{\eta}) - \log \mathcal{L}(f_{\rm sig}, \eta) \right]$ with two degrees of freedom.

The code to perform the analysis has been uploaded to GitHub~\cite{living}. It computes the test statistic and $p$-value for the observed events and scrambled samples, and generates signal injections if a dedicated function is implemented. The repository contains example scripts to reproduce the results presented in the following, and the code can be adapted to any other neutrino datasets and GRB catalogs.

\section{Results}
\label{sec:results}

\begin{figure}[hbtp]
    \centering
    \includegraphics[width=0.5\linewidth]{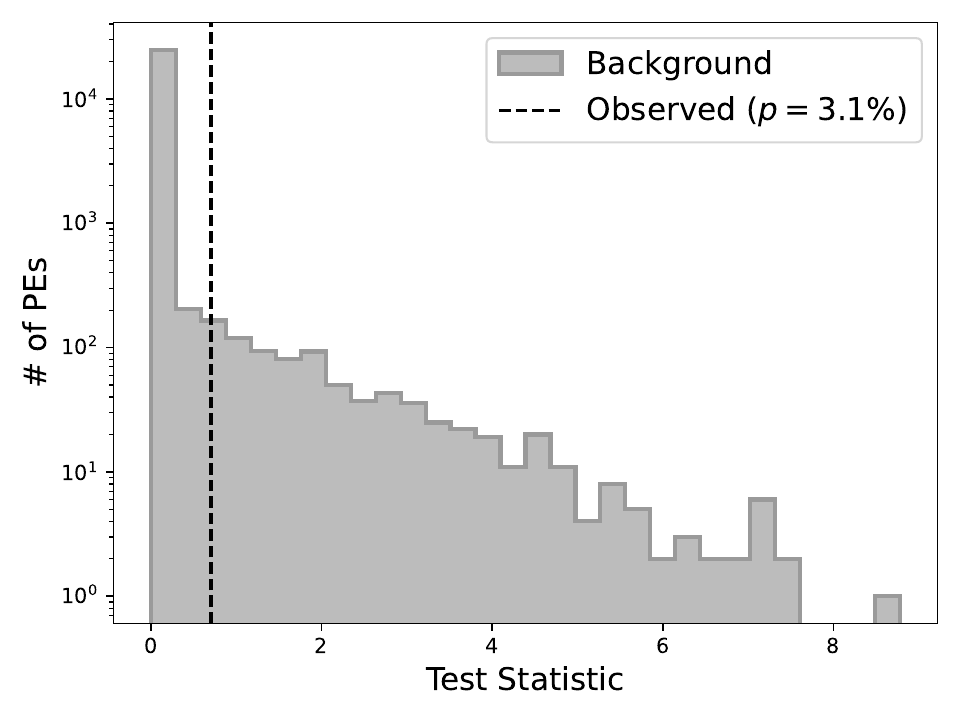}
    \caption{Distribution of the test statistic obtained from scrambled data. The black vertical line is the test statistic value obtained with the unscrambled IceCube HESE 12-year dataset.}
    \label{fig:ts}
\end{figure}

\begin{figure}[hbtp]
    \centering
    \includegraphics[width=0.7\linewidth]{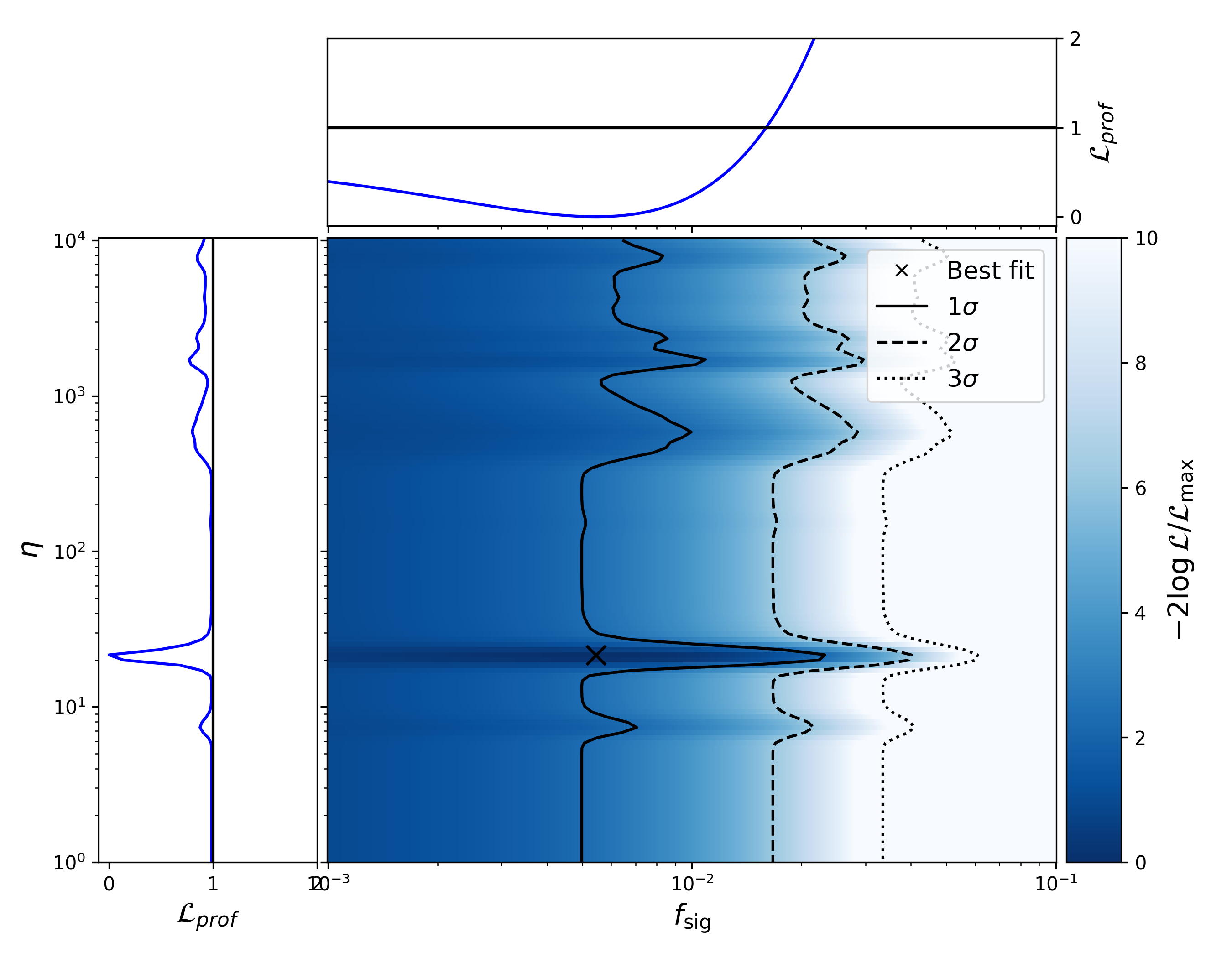}
    \caption{Likelihood distribution for the IceCube HESE 12-year dataset and corresponding 2D contours (at $1-2-3\sigma$ confidence level) using Wilks' theorem. The 1D plots on the top and left show the profiled likelihood on $f_{\rm sig}$ and $\eta$, respectively.}
    \label{fig:contour}
\end{figure}

In the case of first-order LIV ($n = 1$), the observed test-statistic value is $\mathcal{T}_{\rm obs} \simeq 0.7$, leading to a $p$-value of $p=3.1\%$, as illustrated in \autoref{fig:ts}. The best-fit point is $(\hat{f}_{\rm sig}, \hat\eta) = (\num{6.1e-3}, 21.0)$ and the $1-2-3\sigma$ contours are shown in \autoref{fig:contour}. If $\eta$ is set to $0$ and only $f_{\rm sig}$ is fitted, corresponding to the case where there are neutrinos from GRBs but no LIV, the best fit is $f_{\rm sig} = 0$.

\begin{figure}[hbtp]
    \centering
    \includegraphics[width=0.85\linewidth]{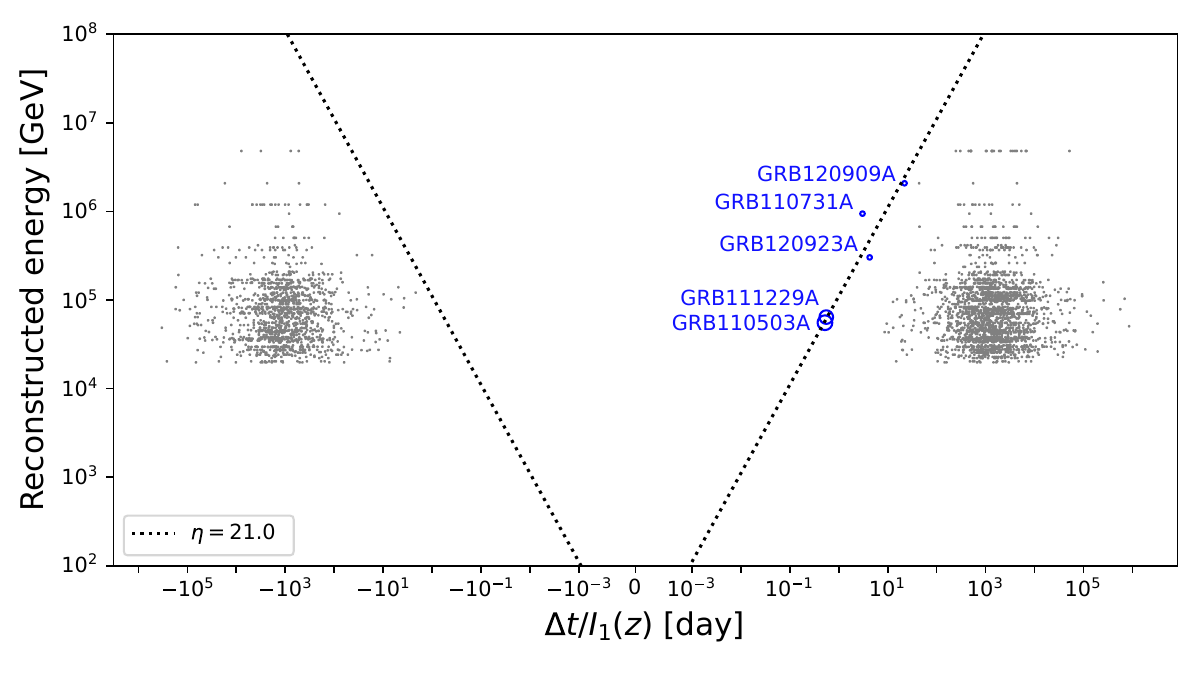}
    \caption{Event distribution in the $E_{\rm rec} - \Delta t/I_1(z)$ plane for the IceCube HESE 12-year dataset. For each neutrino event, we represent only the GRBs at a distance of at most $\qty{20}{\degree}$. The markers in blue (grey) represent the associations with non-zero (zero) signal PDF, and the size of each blue marker is proportional to the value of the signal PDF term. The labels indicate the name of the corresponding GRB. The dotted line shows the expectation for the signal assuming LIV with $\eta = 21.0$.}
    \label{fig:deltaT-E}
\end{figure}

\autoref{fig:deltaT-E} shows the distribution of all associations in the reconstructed energy vs $\Delta t / I_1(z)$ plane, where signal events are expected to align along a line with slope $H_0 \times M_P / \eta$ following \autoref{eq:time-delay2}, neglecting the intrinsic time delay. The top five associations (with the highest $S_{i,j}(\hat\eta)$) are given in \autoref{tab:best-assoc}.

\begin{table}[hbtp]
    \centering
    \begin{tabular}{c|c|ccc}
        \toprule
        \# & Associated GRB & {$E_{\rm rec}$ [TeV]} & {$\Delta t$ [d]} & {$\Delta t_{\rm LIV}^{\eta=21.0}$ [d]} \\
        \midrule
        1 & GRB110503A & 55 & 0.93 & 0.89 \\
        2 & GRB111229A & 64 & 0.85 & 0.89 \\
        3 & GRB120909A & 2075 & 86 & 73 \\
        4 & GRB120923A & 302 & 28 & 18 \\
        5 & GRB110731A & 943 & 9 & 25 \\
        \bottomrule
    \end{tabular}
    \caption{Top four neutrino-GRB associations based on the value of the signal term $S_{ij}$ computed for the best-fit $\eta=21.0$, assuming first-order LIV. The columns display the ranking, the name of the GRB (indicating also its date), the neutrino reconstructed energy, the observed time delay, and the expected time delay for $\eta=21.0$.}
    \label{tab:best-assoc}
\end{table}

The situation changes qualitatively if neutrino decay is neglected by setting $\mathcal{P}_{\rm surv} = 1$, so that both subluminal and superluminal components can reach the detector without attenuation. In this case, we obtain a best-fit result fully consistent with the background-only hypothesis: $p = 100\%$, $(\hat{f}_{\rm sig}, \hat\eta) = (0, 0)$. Therefore, accounting for neutrino decay is essential for the LIV interpretation of the result.

At the opposite extreme, if the analysis is repeated by setting $\mathcal{P}_{\rm surv} = 0$, corresponding to the limiting case of maximal decay in which the superluminal component is completely suppressed, the results remain essentially unchanged, with the best-fit point and the $p$-value shifting by less than the quoted significant digits. This is expected since, for $\eta \sim 10-100$, superluminal neutrinos are strongly suppressed by the decay above \qty{10}{\tera\eV}, while the IceCube HESE sample starts at higher energies. Thus, the main result is effectively driven by the subluminal component and can be reproduced in the limit in which only delayed neutrinos reach the detector. However, consistency with the $n=1$ EFT scenario requires the inclusion of the accompanying superluminal component, whose absence at HESE energies is naturally accounted for by LIV-induced neutrino decays.

These results may be compared with the analysis in \cite{Amelino-Camelia:2025lqn}, reporting a $p$-value $p=0.6\%$ and a best-fit $\eta=21.7$. While superluminal neutrinos were not included in that work as an ad hoc assumption, their effective suppression at HESE energies helps explain the similar best-fit LIV scale. The four most significant neutrino-GRB associations also match those reported in \autoref{tab:best-assoc}. This agreement is found despite the different statistical approach adopted in that work. In particular, their analysis is not based on a likelihood approach and assumes a simpler treatment of the spatial association between neutrinos and GRBs; the correlation is measured by identifying points aligned in the $\Delta t-E$ distribution, neglecting the neutrino energy resolution. Additionally, the GRB catalogs are different (652 sources in \cite{Amelino-Camelia:2025lqn}, compared to 746 in this work), though this leads to only marginal changes when repeating our analysis with their catalog ($p=2.9\%$, $\hat{f}_{\rm sig} = \num{6.5e-3}$, $\hat\eta = 21.0$).

Finally, the constraints on $f_{\rm sig}$ may be converted into ones on the total per-flavour neutrino+antineutrino flux from GRBs ${\rm d}N/{\rm d}E = c \times (E/\unit{\GeV})^{-\gamma}$, accounting for the IceCube HESE effective area~\cite{IceCube:2020wum}, or on the contribution of such sources to the diffuse flux $\Phi(E) = \phi \times (E/\unit{\GeV})^{-\gamma}$. Averaging over all the sky, one gets:
\begin{equation}
    c = \dfrac{f_{\rm sig} \times N_{\nu}}{\displaystyle\int A_{\rm eff}(E) \left(\dfrac{E}{\unit{\GeV}}\right)^{-\gamma} {\rm d}E}
\end{equation}
and
\begin{equation}
    \phi = \dfrac{c \times R_{\rm GRBs}}{4\pi \times N_{\rm GRBs}},
\end{equation}
where $R_{\rm GRBs}$ ($N_{\rm GRBs}$) is the rate (number) of GRBs. The best-fit value $\hat{f}_{\rm sig} = \num{6.1e-3}$ corresponds to an $E^{-2}$ ($E^{-2.52}$, from IceCube combined fit~\cite{IceCube:2025tgp}) flux with $c = \qty{0.21}{\per\GeV\per\square\cm}$ ($\qty{1.5e2}{\per\GeV\per\square\cm}$). Considering $R_{\rm GRBs}$ to be the rate of \emph{observed} GRBs (independently from the availability of a redshift measurement), using data from 2015--2025 in GRBweb~\cite{GRBweb}, one gets $R_{\rm GRBs} = \qty{0.84}{\per\day}$ and $\phi = \qty{2.2e-10}{\per\GeV\per\square\cm\per\s\per\steradian}$ ($\qty{1.5e-7}{\per\GeV\per\square\cm\per\s\per\steradian}$) for $\gamma = 2$ ($\gamma = 2.52)$.

The full analysis was repeated for second-order LIV, showing no evidence for any deviation from the background-only hypothesis: $p = 100\%$, $(\hat{f}_{\rm sig}, \hat\eta) = (0, 0)$.

\section{Discussion and outlooks}
\label{sec:discussion}

As highlighted, measuring a redshift- and energy-dependent time delay between photons and neutrinos from gamma-ray bursts would constitute solid evidence for processes beyond the Standard Model, such as LIV. Using one of the most recent public IceCube datasets and the GRBweb catalog, the search for such delays shows a slight hint for the presence of LIV of order $n=1$, with a $p$-value of $3.1\%$, corresponding to a one-sided Gaussian significance of approximately $1.9\sigma$. The best fit is found at a scale $\Lambda_{\rm LIV} = M_P/\eta = \qty{5.8e17}{\GeV}$.

The constraints on the corresponding neutrino flux should be considered lightly, as the analysis assumes equal weighting for all the GRBs, therefore not accounting for distance and source-by-source variations of the expected neutrino signal. A more detailed study would require the inclusion of proper models predicting a specific spectrum for each GRB. Within these limitations, assuming a spectral index $\gamma = 2.52$, the best fit corresponds to a diffuse neutrino flux from GRBs $\left.E^2 \Phi\right\vert_{\qty{1}{\GeV}} = \qty{1.5e-7}{\GeV\per\square\cm\per\s\per\steradian}$ (per-flavour, $\nu+\bar\nu$), which is about 2\% of the total diffuse flux~\cite{IceCube:2025tgp}.

The presented work relies on the IceCube HESE 12-year public dataset, available with only limited information on the detector response, especially for the energy resolution. The use of more detailed responses and the inclusion of additional neutrino samples with complementary performance (for instance, cascade events) would allow confirming the result, eventually improving its significance and the exclusion contours shown in \autoref{fig:contour}.

It is worth noting that the results would not change if the superluminal component were completely suppressed, while the excess is no longer significant if decay is disabled. The former point is due to the strong suppression of superluminal neutrinos for the relevant range of LIV scale and neutrino energies considered here. At lower energies, however, the superluminal component would no longer be completely suppressed, making neutrino samples extending to the \unit{\TeV} range particularly relevant.

Interestingly, the limiting case in which only the subluminal branch contributes admits an alternative interpretation beyond the EFT scenario considered here. A universal subluminal LIV scenario, in which the same modification applies to all particle species and equally to particles and antiparticles, would also produce only the delayed branch, without requiring the suppression of an accompanying superluminal component. Since the photons used to identify GRBs have energies much lower than those of the neutrinos considered here, their LIV-induced delays would be negligible, leaving the relative neutrino–photon time delay effectively unchanged. At the level of the present analysis, this scenario is therefore phenomenologically indistinguishable from the limit $\mathcal P_{\rm surv}=0$, although the underlying physics is different: in the EFT scenario the superluminal component is present but strongly suppressed by neutrino decays, whereas in the universal scenario it is absent from the outset. Such a universal modification would also alleviate the constraints associated with electron stability recently highlighted in Ref.~\cite{Bustamante:2026rff}. TeV-scale neutrino samples could therefore provide a way to distinguish these two interpretations through the possible reappearance of the superluminal component expected in the EFT scenario.

The obtained results are aligned with similar studies in the literature \cite{Amelino-Camelia:2022pja,Amelino-Camelia:2025kro,Zhang:2022svg}. We plan to apply the same methodology with other datasets, such as the ones from ANTARES~\cite{ANTARES:2011hfw} and KM3NeT~\cite{KM3Net:2016zxf}, especially re-evaluating the contribution of KM3-230213A~\cite{KM3NeT:2025npi}, claimed to be potentially associated with GRB090401B~\cite{Amelino-Camelia:2025lqn,Wang:2026eyt}.

Such further studies are clearly required, as the evidence is yet insufficient, though the existence of LIV would have a strong impact, both in the landscape of Standard Model extensions and in the study of transient astrophysical sources.

\acknowledgments{ML is funded by the European Union’s Horizon Europe Research and innovation programme under Grant Agreement No 101131928. AG is supported by the predoctoral contract CPE\_27\_25 funded by the Gobierno de Aragón under the 2025–2029 programme for predoctoral research personnel. MAR is supported by Project CAPICLE (PDIP-2026-1575) funded by the Croatian Science Foundation (HrZZ). We acknowledge support by the Spanish grant PID2024-160228NB-I00, funded by the Agencia Estatal de Investigación (MCIN/AEI/10.13039/501100011033) and ERDF -- A way of making Europe, and by the Gobierno de Aragón through the research group E21\_26R (Grupo Teórico de Física de Altas Energías), as well as the contribution of the COST Actions CA18108 ``Quantum gravity phenomenology in the multi-messenger approach'' and CA23130 ``Bridging high and low energies in search of quantum gravity''. Finally, we gratefully acknowledge support from the CNRS/IN2P3 Computing Center (Lyon - France) for providing computing and data-processing resources needed for this work.}

\bibliography{references}

\end{document}